\documentclass[prd,aps,showpacs,reprint,nofootinbib,floatfix,superscriptaddress]{revtex4-1}
\usepackage[utf8]{inputenc}
\usepackage{graphicx}
\usepackage{epsfig}
\usepackage{rotating}
\usepackage{amssymb}
\usepackage{subfigure}
\usepackage{dsfont}
\usepackage{psfrag}
\usepackage{amsmath,euscript,array,mathrsfs}
\usepackage{bbold}
\usepackage{epsf}
\usepackage{axodraw}
\usepackage{hyperref}

\usepackage[T1]{fontenc} 

\newcommand{\beq}{\begin{equation}}
\newcommand{\eeq}{\end{equation}}
\newcommand{\beqs}{\begin{eqnarray}}
\newcommand{\eeqs}{\end{eqnarray}}
\newcommand{\lsim}{\mathrel{\raisebox{-
.6ex}{$\stackrel{\textstyle<}{\sim}$}}}

\newcommand{\Tr}{{\rm Tr}}

\newcommand\SU{\mathrm{SU}}

\def\hbar{\hspace{0pt}\raisebox{1pt}{$-$} \hspace{-7pt} h}

\newcommand{\be}{\begin{equation}}
\newcommand{\ee}{\end{equation}}

\newcommand{\bea}{\begin{eqnarray}}
\newcommand{\eea}{\end{eqnarray}}

\def\lbldef#1#2{\expandafter\gdef\csname #1\endcsname {#2}}

\def\href#1#2{#2}

\newcommand{\ber}{\begin{eqnarray}}
\newcommand{\eer}{\end{eqnarray}}

\newcommand{\beqar}{\begin{eqnarray}}

\newcommand{\eeqar}{\end{eqnarray}}

\newcommand{\dsl}
  {\kern.06em\hbox{\raise.15ex\hbox{$/$}\kern-.56em\hbox{$\partial$}}}

\newcommand{\eeqarr}{\end{eqnarray}}
\newcommand{\ZZ}{{\rm \kern 0.275em Z \kern -0.92em Z}\;}

\def\CC{{\mathchoice
{\rm C\mkern-8mu\vrule height1.45ex depth-.05ex
width.05em\mkern9mu\kern-.05em}
{\rm C\mkern-8mu\vrule height1.45ex depth-.05ex
width.05em\mkern9mu\kern-.05em}
{\rm C\mkern-8mu\vrule height1ex depth-.07ex
width.035em\mkern9mu\kern-.035em}
{\rm C\mkern-8mu\vrule height.65ex depth-.1ex
width.025em\mkern8mu\kern-.025em}}}

\def\RR{{\rm I\kern-1.6pt {\rm R}}}

\def\ZZ{{\rm Z}\kern-3.8pt {\rm Z} \kern2pt}
\def\IB{\relax{\rm I\kern-.18em B}}
\def\ID{\relax{\rm I\kern-.18em D}}
\def\II{\relax{\rm I\kern-.18em I}}
\def\IP{\relax{\rm I\kern-.18em P}}

\newcommand{\bear}{\begin{eqnarray}}
\newcommand{\eear}{\end{eqnarray}}

\def\6{\partial}

\def\bea{\begin{eqnarray}}
\def\eea{\end{eqnarray}}

\def\beqx{\begin{displaymath}}
\def\eeqx{\end{displaymath}}

\newcommand{\bmat}{\left(\begin{array}}
\newcommand{\emat}{\end{array}\right)}

\def\bo{{\raise-.3ex\hbox{\large$\Box$}}}               
\def\face{{\raise.2ex\hbox{$\displaystyle \bigodot$}\mskip-2.2mu \llap {$\ddot
        \smile$}}}                                   
\def\>{\rangle}                                      
\def\<{\langle}                                      

\def\leftrightarrowfill{$\mathsurround=0pt \mathord\leftarrow \mkern-6mu
        \cleaders\hbox{$\mkern-2mu \mathord- \mkern-2mu$}\hfill
        \mkern-6mu \mathord\rightarrow$}        
\def\dvec#1{\vbox{\ialign{##\crcr
        \leftrightarrowfill\crcr\noalign{\kern-1pt\nointerlineskip}
        $\hfil\displaystyle{#1}\hfil$\crcr}}}           
\def\Tr{{\rm Tr \,}}                                    

\def\-{\hphantom{-}}

\begin{document}

\title{The Dilaton Potential and Lattice Data}

\author{Thomas Appelquist}
\affiliation{Department of Physics, Sloane Laboratory, Yale University, Prospect Street, New Haven, Connecticut 06520, USA}
\author{James Ingoldby}
\affiliation{Department of Physics, Sloane Laboratory, Yale University, Prospect Street, New Haven, Connecticut 06520, USA}
\affiliation{Abdus Salam International Centre for Theoretical Physics, Strada Costiera 11, 34151, Trieste, Italy}
\author{Maurizio Piai}
\affiliation{Department of Physics, College of Science, Swansea University, Singleton Park, SA2 8PP, Swansea, Wales, UK}


\begin{abstract}
We study an effective field theory (EFT) describing the interaction of an 
approximate dilaton with a set of pseudo-Nambu-Goldstone bosons (pNGBs). The EFT 
is inspired by, and employed to analyse, recent results from lattice calculations that reveal the presence of a remarkably light singlet scalar particle. We adopt a simple form for the scalar potential for the EFT, which interpolates among earlier proposals. It describes departures from conformal symmetry, by the insertion of a single operator at leading order in the EFT. To fit the lattice results, the global internal symmetry is explicitly broken, producing a common mass for the 
pNGBs, as well as a further, additive deformation of the scalar potential. 
We discuss sub-leading corrections arising in the EFT from quantum loops.  
From lattice measurements of the scalar and pNGB masses and of the pNGB decay constant, we extract model parameter values, including those that characterise the scalar potential. The result includes the possibility that the conformal 
deformation is clearly non-marginal. The extrapolated values for the decay constants and the scalar mass would then be not far below the current lattice-determined values.
\end{abstract}

\maketitle

\section{Introduction}
\label{Sec:Introduction}

Lattice studies of $SU(3)$ gauge theories with matter field content consisting of 
$N_f=8$ fundamental (Dirac) fermions~\cite{Appelquist:2016viq,
Gasbarro:2017fmi,Appelquist:2018yqe,Aoki:2014oha,Aoki:2016wnc},
as well as $N_f=2$ symmetric 2-index (Dirac) fermions (sextets)~\cite{Fodor:2012ty,Fodor:2015vwa,Fodor:2016pls,
Fodor:2017nlp,Fodor:2019vmw}, have reported  evidence of the presence in the spectrum of a light scalar, singlet particle,
 at least  in the accessible range of fermion masses.
Motivated by the possibility that such a particle might be 
 a dilaton, the  scalar particle associated with
the spontaneous breaking of scale invariance, in Refs.~\cite{Appelquist:2017wcg,Appelquist:2017vyy} 
we analysed lattice data in terms of
 an effective field theory (EFT) framework that extends the field content 
 of the chiral Lagrangian. It includes a dilaton field $\chi$, 
together with the pseudo-Nambu-Goldstone-boson (pNGB) fields $\pi$, along the lines discussed also in Refs.~\cite{Golterman:2016lsd,Golterman:2016hlz,Golterman:2016cdd,Golterman:2018mfm} and more generally in Refs.~\cite{Goldberger:2008zz,Matsuzaki:2013eva,Kasai:2016ifi,Hansen:2016fri}.

Dilaton EFTs (see Ref.~\cite{Coleman:1985rnk} for a pedagogical introduction) are of general interest, reaching well beyond the study of $SU(3)$ lattice gauge theories. The recently discovered Higgs particle~\cite{Aad:2012tfa,Chatrchyan:2012xdj} might originate as a light dilaton in  extensions of the standard model (SM) of particle physics. Suggestions in this direction can be found for example in Refs.~\cite{Leung:1985sn,Bardeen:1985sm,Yamawaki:1985zg}, and more recently these ideas have been revived in the context of compositeness and dynamical electroweak symmetry breaking (see for instance Refs.~\cite{Hong:2004td,Dietrich:2005jn,Hashimoto:2010nw,Appelquist:2010gy,Vecchi:2010gj,Chacko:2012sy,Bellazzini:2012vz,Abe:2012eu,Eichten:2012qb,Hernandez-Leon:2017kea}).

The Lagrangian density is comparatively simple.
Besides the kinetic terms, it  contains two other
types of terms, responsible for explicit breaking of scale invariance: a scalar potential $V(\chi)$ for the dilaton field,
and an additional operator that depends both on the pNGB and dilaton fields, which generates masses for the pNGBs.
The latter must be included because the lattice formulation of the gauge theories of interest requires a mass term for the fermions,
which explicitly breaks both scale invariance as well as chiral symmetry.

By studying the pNGB mass and decay constant as a function of the 
fermion mass, one can extract from the lattice data for the two  
nearly-conformal gauge theories information about the potential
$V(\chi)$ and other dynamical quantities. An example is a 
certain parameter $y$, interpreted as the mass-dimension of the chiral 
condensate in the underlying theory in its strong-coupling regime~\cite{Leung:1989hw}. In Ref.~\cite{Appelquist:2017wcg} we found, independently of $V(\chi)$, $y\sim 2$, compatible with expectations from studies of strongly-coupled gauge theories near the edge of the conformal window~\cite{Cohen:1988sq}
(see also Refs.~\cite{Ryttov:2017kmx} and~\cite{Aoki:2016wnc}). 
We also found the shape of the potential at large field excursions to be compatible 
with a simple power-law, $V(\chi) \propto \chi^p$ with $p$ close to 4~\cite{Appelquist:2017vyy}. By combining this result with the lattice measurement of the scalar mass, we estimated the ratio of decay constants of the dilaton 
and the pNGBs, finding roughly $f_{\pi}^2 /f_d^2 \sim 0.1$. For both of the 
two gauge theories, the dilaton EFT fared remarkably well 
when used at the tree-level.

In this paper, we address open questions facing the dilaton EFT 
framework, and further assess its potential as a tool in the 
interpretation of lattice data, here for the $N_f=8$, $SU(3)$ gauge theory. To study extrapolation from the regime of lattice data to the limit of massless 
NGBs, we adopt a specific form for the scalar potential $V(\chi)$. In addition 
to the scale invariant term $\sim\chi^4$, we include a single operator with 
generic scaling dimension to break the conformal 
symmetry~\cite{Goldberger:2008zz,Golterman:2016lsd,Coleman:1985rnk}. We find that 
the lattice data allows for the conformal deformation to be clearly 
non-marginal. 

We discuss sub-leading corrections arising from quantum loops 
(see also Refs.~\cite{Kasai:2016ifi,Hansen:2016fri,Bijnens:2009qm}) 
and new operators within the EFT. We display the new 
operators that correct the potential in the extrapolated chiral limit, 
and that are generated in a loop expansion. 
We argue that corrections to the dilaton EFT are parametrically suppressed in this limit, and lead to relatively small effects. At finite fermion mass, in the regime of the lattice data, we examine the class of loop corrections corresponding to distortions of the tree-level scalar potential, finding that they are small.

Section~\ref{Sec:Lagrangian} introduces the dilaton 
EFT. We display all the tree-level scaling relations used in the 
analysis of the lattice data. In Section~\ref{Sec:Lattice} we perform a numerical global 
fit of the lattice data taken from Ref.~\cite{Appelquist:2018yqe} for the $N_f = 8$, $SU(3)$ gauge 
theory, determining the six independent parameters of the tree-level EFT.
In Section~\ref{Sec:Corrections} we discuss sub-leading corrections and estimate their size.
We conclude by summarising and discussing our main findings in Section~\ref{Sec:Discussion}.

\section{Lagrangian density and scaling relations}
\label{Sec:Lagrangian}

We develop further the framework we adopted  in Refs.~\cite{Appelquist:2017wcg,Appelquist:2017vyy},
by assuming that the strongly-coupled dynamics of the underlying nearly conformal gauge theory is captured by a dilaton EFT satisfying the following conditions.

\begin{itemize}

\item[1.] The EFT is governed by scale invariance over a finite range of scales. The long distance dynamics yields a space of degenerate, inequivalent vacua, which are parametrised by a scalar field $\chi$. Scale invariance is spontaneously broken through a 
non-vanishing vacuum value $f_d$ for $\chi$. A massless scalar particle, the
dilaton, then appears, with its couplings set by $f_d$.

\item[2.] Scale invariance is explicitly broken, while keeping $f_d$ fixed. The dilaton is given a small mass $m_d$. As a consequence, this explicit breaking is suppressed.

\item[3.] The EFT  admits an internal {\bf continuous global symmetry}
with Lie group $G$, broken spontaneously  to a subgroup $H$.
The spectrum includes a set of pNGBs with couplings set by their decay constant $f_{\pi}$.

\item[4.] The global symmetry is $G$ is then broken by introducing a {\bf tunable mass} $m_\pi^2$ for the pNGBs associated with the spontaneous breaking of $G$. This mass also breaks scale invariance explicitly, through an operator with {\bf scaling dimension $y$}. The vacuum of the theory is shifted, and we denote by $F_{\pi}$, $F_d$ and $M_{\pi}$ the decay constants of the pNGBs and dilaton, and the pNGB mass in this vacuum.

\item[5.]  In the limit $m^2_\pi \rightarrow 0$, the vacuum of the theory is selected dominantly by a single operator in the EFT with {\bf scaling dimension} $\Delta$. 
When $m^2_\pi \neq 0$, there are two sources of dilaton mass, and we denote the full mass by $M_d$.

\end{itemize}

In  Refs.~\cite{Appelquist:2017wcg,Appelquist:2017vyy} we made use of conditions 1 through 4, with $G=SU(N_f)_L\times SU(N_f)_R$ and  $H=SU(N_f)_V$. We studied  an EFT in which the pNGB and dilaton particles are described by a set of fields $\pi$  in the coset $G/H$ and an additional real scalar field $\chi$. The tree-level Lagrangian density is then the following~\cite{Appelquist:2017vyy}:
\beqs
{\cal L}&=&\frac{1}{2}\partial_{\mu}\chi\partial^{\mu}\chi\,+{\cal L}_{\pi} \,+\,{\cal L}_M\,-\,V(\chi)\,,
\label{Eq:L}
\eeqs
where the dynamics of the pNGBs is governed by
\beqs
{\cal L}_{\pi}&=&\frac{f_{\pi}^2}{4}\left(\frac{\chi}{f_d}\right)^2 \,{\rm Tr}\,\left[\partial_{\mu}\Sigma\frac{}{}(\partial^{\mu}\Sigma)^{\dagger}\right]\,.
\label{Eq:Lpi}
\eeqs 
The matrix-valued field  $\Sigma=\exp\left[2i\pi/f_\pi\right]$ transforms as $\Sigma\rightarrow U_L \Sigma U_R^{\dagger}$ under the action of unitary transformations $U_{L,R}\in SU(N_f)_{L,R}$.
It also satisfies the non-linear constraints $\Sigma \Sigma^{\dagger}=\mathbb{1}_{N_f}$.
The explicit breaking of the global internal symmetry is captured in the EFT by the term
\beqs
{\cal L}_M&=&\frac{m_{\pi}^2f_{\pi}^2}{4}\left(\frac{\chi}{f_d}\right)^y \,{\rm Tr}\,\left[\Sigma\frac{}{}+\frac{}{}\Sigma^{\dagger}\right]\,,
\label{Eq:LM}
\eeqs
where $m_{\pi}^2\equiv 2 B_{\pi}m$ vanishes when the fermion mass $m$ of the underlying theory is set to zero. An interpretation of $y$ as the scaling dimension of the chiral condensate of the underlying gauge theory at strong coupling can be found for example in Ref.~\cite{Leung:1989hw}.


The Lagrangian must contain an additional, explicit source of breaking of scale invariance, in the potential $V(\chi)$.
In Refs.~\cite{Appelquist:2017wcg,Appelquist:2017vyy}, we showed that one can in principle reconstruct the functional dependence of $V(\chi)$ 
indirectly, by studying the dependence on the fermion mass of appropriate combinations of the mass and decay constant of the pNGBs. To make further progress, we now commit to a specific class of potentials. We make use of condition 5, employing a single operator breaking the scale invariance of the potential.

We adopt the following choice of tree-level potential $V(\chi)$:
\beqs
V_{\Delta}(\chi)&\equiv&\frac{m_d^2\chi^4}{4(4-\Delta)f_d^2}\left[1-\frac{4}{\Delta}\left(\frac{\chi}{f_d}\right)^{\Delta-4} \right]\,.
\label{Eq:potential}
\eeqs
This potential, also discussed in Refs.~\cite{Chacko:2012sy,Cata:2018wzl}, contains two contributions. One is a scale-invariant term ($ \propto \chi^4$) representing the corresponding operators in the underlying
gauge theory. The other ($ \propto \chi^{\Delta}$) captures the leading-order effect of the scale 
deformation in the underlying theory. 
The normalisation of the two coefficients is such that the potential has 
a minimum at $\chi = f_d  > 0$, with curvature $m_d^2$, corresponding to mass $m_d$ for the dilaton.

In Section \ref{Sec:Lattice}, we employ $V_{\Delta}$ to fit lattice data on the $\SU(3)$ theory with $N_f=8$ Dirac fundamental fermions. We then return to Eq.~(\ref{Eq:L}) and to $V_{\Delta}$ in Section \ref{Sec:Corrections}, to show explicitly that this is the leading-order part of a systematic expansion for the EFT in a small parameter. We perform a spurion analysis for the potential terms, which we relegate to Appendix \ref{Sec:appendixa}. Then in Section \ref{Sec:Corrections}, we classify and estimate the magnitude of corrections to the leading-order Lagrangian, by means of a perturbative loop analysis.

A key feature of the EFT is that the dilaton mass (the explicit breaking of scale symmetry) can be tuned as small as necessary with $f_d$ held fixed. In the limit, the space of VEVs becomes a moduli space, presumably reflecting the same feature of the underlying gauge theory. It has been suggested in Refs.~\cite{Appelquist:2010gy} and \cite{Golterman:2016lsd} that the explicit breaking in the underlying theory, and also in the EFT as a consequence,  can be made arbitrarily small by tuning the number of flavours $N_f$ arbitrarily close the the critical value $N_{f}^{c}$ at which confinement gives way to IR conformality, with the emergence of  fixed points generalising Refs.~\cite{Caswell:1974gg,Banks:1981nn}. This can be arranged by taking $N_f$ to be a continuous parameter or working in the large-$N$ limit. The authors of Ref.~\cite{Golterman:2016lsd} make extensive use of this plausible idea. We instead work only with the EFT, employing condition 1 as one of its principles.

We note finally that the form of $V_{\Delta}(\chi)$ interpolates among several specific forms found in the
 literature. In Ref.~\cite{Appelquist:2017wcg}, we fitted the lattice data available then for the $N_f = 8$ 
 theory employing two forms of some historical interest. 
 The choice $\Delta = 2$ gives the Higgs potential of the standard model (up to an inconsequential additive constant) 
\beqs
V_1&\equiv&\frac{m_d^2}{2f_d^2}\left(\frac{\chi^2}{2}-\frac{f_d^2}{2}\right)^2\,.
\label{Eq:V1}
\eeqs
The choice $\Delta \rightarrow 4$, corresponding to a marginal deformation of scale symmetry, leads to
\beqs
V_2&\equiv&\frac{m_d^2}{16f_d^2}\chi^4\left(4\ln\frac{\chi}{f_d}-1\right)\,.
\label{Eq:V2}
\eeqs
Discussion of this form can be found in Refs.~\cite{Bardeen:1985sm} and~\cite{Migdal:1982jp}. 
It is also considered in Ref.~\cite{Goldberger:2008zz}.
 Another form~\cite{Coleman:1985rnk}, illustrating the principles for building a dilaton 
 potential, corresponds to the limiting case $\Delta \rightarrow 0$.

\subsection{Scaling relations}
\label{Sec:Scaling}

Here we summarise properties of the EFT and its predictions to be used to study the numerical lattice data. We 
draw on Refs.~\cite{Appelquist:2017vyy,Appelquist:2017wcg} supplemented by explicit use of the potential $V_{\Delta}$ in Eq.~(\ref{Eq:potential}). The mass deformation encoded in Eq.~(\ref{Eq:LM}) contributes, in the vacuum $\langle\pi\rangle = 0$, an additive term to $V_{\Delta}$. The entire potential is 
\beqs
W(\chi)&=& V_{\Delta}(\chi) \,-\,\frac{N_f m_{\pi}^2f_{\pi}^2}{2}\left(\frac{\chi}{f_d}\right)^y\,,
\label{Eq:Wpotential}
\eeqs 
leading to a new minimum for $\chi$ which determines its vacuum value $\langle\chi\rangle = F_d > f_d$. Also, there is a new curvature at this minimum, determining the dilaton mass $M_d^2$. 

By employing the value $\langle \chi \rangle = F_d$ in Eqs.~(\ref{Eq:Lpi}) and (\ref{Eq:LM}), and properly normalising the 
pNGB kinetic term, the simple scaling relations for the 
pNGB decay constant and mass derived in Ref.~\cite{Appelquist:2017vyy} can be 
found. These relations, which are independent of the explicit form of the potential, are:
\beqs
\frac{F_{\pi}^2}{f_{\pi}^2}&=& \frac{F_{d}^2}{f_{d}^2}\,,\label{Eq:scaling}\\
\frac{M_{\pi}^2}{m_{\pi}^2}&=&\left( \frac{F_{d}^2}{f_{d}^2}\right)^{\frac{y}{2}-1}\label{Eq:Mpiscaling}\,.
\eeqs
The ratio $F_d/f_d$, found by minimising the entire potential in Eq.~(\ref{Eq:Wpotential}), satisfies
\beqs
\left(\frac{F_d}{f_d}\right)^{4-y} \frac{1}{4-\Delta}\left[1 - \left(\frac{f_d}{F_d}\right)^{4-\Delta} \right]  &=& R   \,,
\label{Eq:Fdvariation}
\eeqs
where
\beqs
R \equiv \frac{y N_{f} f_{\pi}^2 m_{\pi}^2 }{ 2 f_{d}^2 m_{d}^2 }.
\label{Eq:R}
\eeqs  
The quantity $m_{\pi}^2$ is in turn related to the fermion mass $m$ in the underlying theory 
by $m_{\pi}^2 = 2  B_{\pi} m$. 
The left-hand side of Eq.~(\ref{Eq:Fdvariation}) is a monotonically increasing function of 
$F_d/f_d$ for any value of $\Delta$ (in the physical region $F_d>f_d$ so long as $y<4$), indicating that $R$ is a 
useful measure of the deformation due to the fermion mass. Large values of $R$ correspond to a large 
deformation, 
with $F_d$ displaced far from its chiral-limit $f_d$. 
 The EFT can be used for only a finite range of fermion mass such that the
 approximate scale invariance of the underlying gauge theory is maintained.
The ratio $M_d^2/ m_d^2$ is given by 
\begin{multline}
\frac{M_d^2}{m_d^2} = \frac{3}{4 - \Delta} \left(\frac{F_d}{f_d}\right)^{2} - 
 \frac{\Delta-1}{4 - \Delta} \left(\frac{F_d}{f_d}\right)^{\Delta-2} \\  - R  (y-1) \left(\frac{F_d}{f_d}\right)^{y-2} \,.
\label{Eq:Mdvariation}
\end{multline}

Eqs.~(\ref{Eq:scaling})-(\ref{Eq:Mdvariation}) can be reorganised into three simple expressions that are more convenient 
for fitting lattice data. Measurements exist only for the quantities $F^2_\pi$, $M^2_\pi$ and $M^2_d$, 
so we eliminate $F^2_d$ from the expressions. First, the two scaling equations (\ref{Eq:scaling}) and (\ref{Eq:Mpiscaling}) can be combined to give
\beqs 
M_{\pi}^2 F_{\pi}^{2 - y} = C \,m \,,
\label{Eq:fit1}
\eeqs
where $C = 2B_\pi f_\pi^{2-y} $ is treated as a fit parameter. Also, it is convenient to combine Eqs.~(\ref{Eq:Fdvariation}) and (\ref{Eq:Mdvariation}) with Eq.~(\ref{Eq:Mpiscaling}), so that the exponential dependence on the unknown fit parameter $y$ is removed from these fit equations. We arrive at 
\beqs
\frac{M_{\pi}^2}{F_{\pi}^2}&=&\frac{2 m_d^2 f_d^2}{y N_f (4-\Delta) f_{\pi}^4}\left(1-\left(\frac{f_{\pi}}{F_{\pi}}\right)^{4 -\Delta}\right)\,,
\label{Eq:fit2}
\eeqs
and 
\beqs
\frac{M_{d}^2}{F_{\pi}^2}&=&\frac{ m_d^2}{ (4-\Delta) f_{\pi}^2}\left(4-y+(y-\Delta) \left(\frac{f_{\pi}}{F_{\pi}}\right)^{4-\Delta}\right).\qquad
\label{Eq:fit3}
\eeqs

\section{Comparison to lattice data}
\label{Sec:Lattice}

We perform a global, six-parameter fit to lattice data, employing Eqs.~(\ref{Eq:fit1})-(\ref{Eq:fit3}). We use the four dimensionless parameters $y$, $\Delta$, $f_{\pi}^2/f_d^2$, and $m_d^2/f_d^2$, along with the two parameters $f^2_{\pi}$ and $C$, expressed in units of the lattice spacing $a$. The larger uncertainty in the measurement of $M_d^2$ limits the precision achievable in extracting certain combinations of these six parameters from the global fit.

\begin{center}
\begin{figure*}
\begin{picture}(500,175)
\put(0,0){\includegraphics[height=5.4cm]{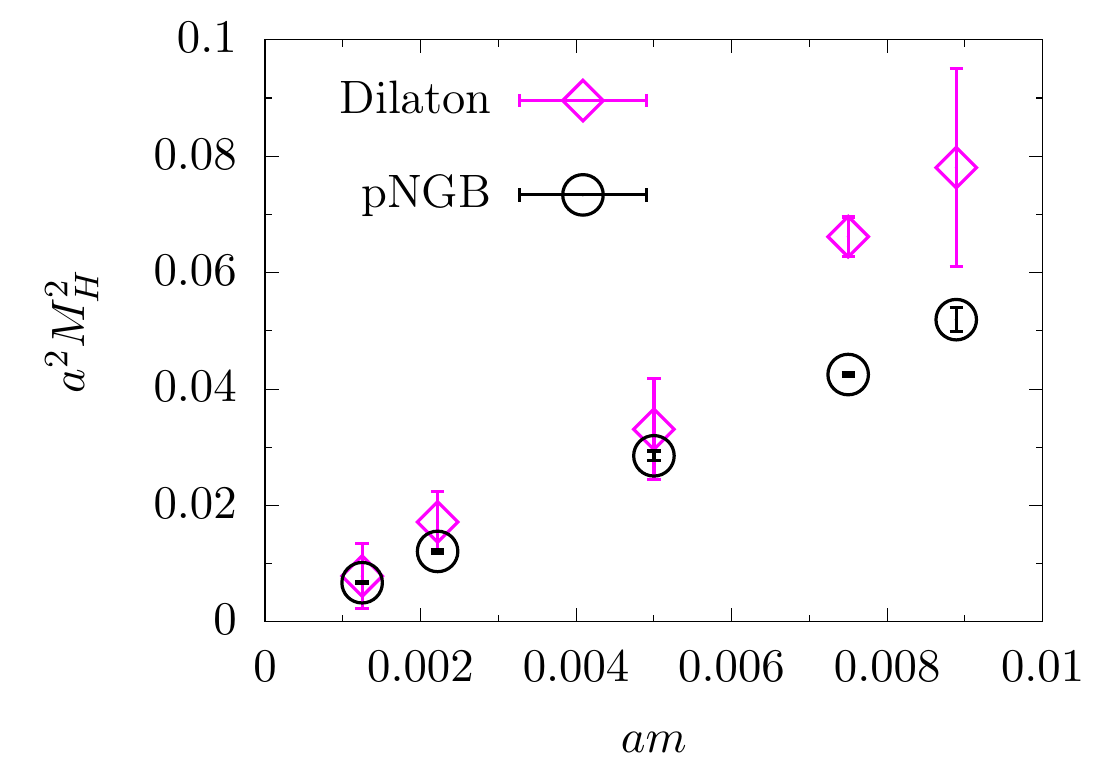}}
\put(260,0){\includegraphics[height=5.4cm]{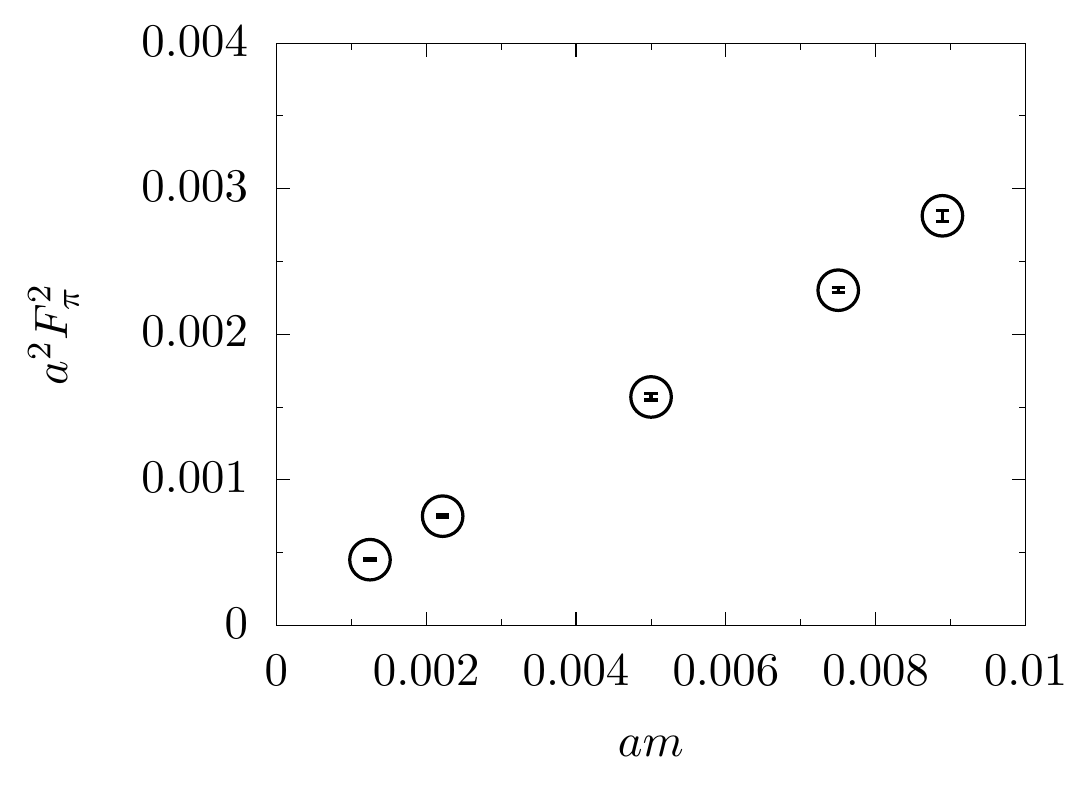}}
\end{picture}
\caption{Lattice measurements for the $SU(3)$ theory with $N_f=8$ fundamentals,
		 obtained from Ref.~\cite{Appelquist:2018yqe}. The lattice spacing is denoted by $a$. For each value of the fermion mass $m$,
		 we show the mass $M_d$ of the dilaton, the  mass $M_{\pi}$ of the pNGBs  and the decay constant $F_{\pi}$ of the pNGBs.	
		 The error bars encompass both statistical and fit-range systematic uncertainties, 
		 as presented in Ref.~\cite{Appelquist:2018yqe}.}
		\label{Fig:dataLSD}		
\end{figure*}
\end{center}

Lattice data for the $SU(3)$ gauge theory with $N_f=8$ Dirac fermions in the fundamental representation
are taken from the tables in Ref.~\cite{Appelquist:2018yqe}. The information
 we use is displayed in Fig.~\ref{Fig:dataLSD}.
The error bars shown on the plots represent combined statistical and fit-range systematic uncertainties, 
but do not include any other systematics, such as lattice artefacts
arising from discretisation and finite volume. We refer to the original publication for details. 
There is no publicly available information about the correlation between
$M_{\pi/d}^2$ and $F_{\pi}^2$, and we therefore treat them as independent measurements. 

\vspace{-30pt}
\begingroup
\renewcommand\arraystretch{1.2}
\setlength{\tabcolsep}{15pt}
\begin{table}[h]	
	\begin{tabular}{ c  c }
		\hline\hline
		Parameter & Central value \\
		\hline
		$y$ & 2.06\\
		$a^{3-y}C$ & 6.9\\
		$\Delta$&  3.5\\
		$a^2 f^2_\pi$ &  $1.2\times10^{-5}$\\
		$f_\pi^2/f_d^2$ &   0.086\\
		$m_d^2 / f_d^2$ &   0.75\\
		\hline\hline
	\end{tabular}
\caption{Central values of (dimensionless) fit parameters obtained in the six-parameter
 fit of Eqs.~(\ref{Eq:fit1})-(\ref{Eq:fit3}) to the LSD data taken from Ref.~\cite{Appelquist:2018yqe}. 
 The uncertainty on these determinations, and
the associated correlations, are discussed in the main text and illustrated 
in Figs.~\ref{Fig:yC}, \ref{Fig:chisquarevsdelta}
and \ref{Fig:fpimdvsdelta}. }
\label{Tab:centralvalues}
\end{table}
\endgroup

Our global fit leads to the set of parameter central values shown in Table~\ref{Tab:centralvalues}.
There are 15 data points in total and 6 fit parameters, yielding $N_\text{dof}=9$.
Evaluated at the minimum, we find $\chi^2/N_\text{dof} = 0.38$. The $\chi^2$ function used in our global fit was constructed by simply summing separate contributions from each of the three fit equations. This function is steeper in some directions in the six parameter space than in others, and there are visible correlations.

\begin{figure}[h]
\begin{center}
\includegraphics[height=5.8cm]{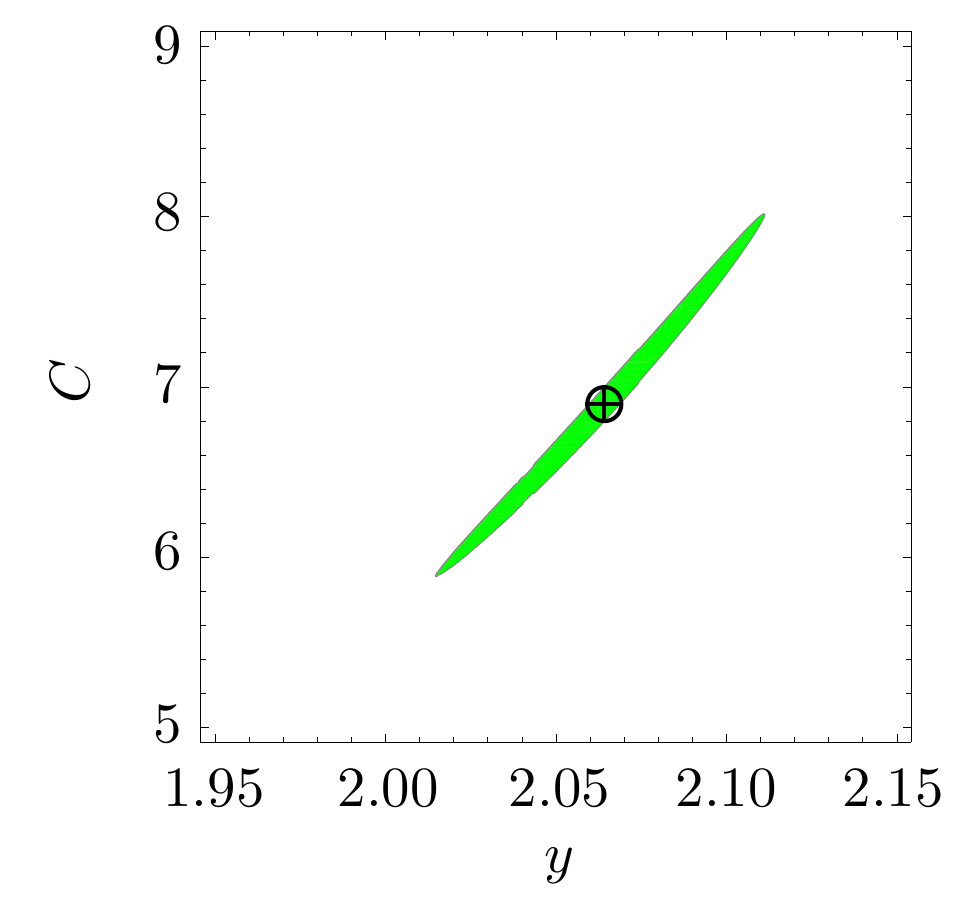}
		\caption{The $1\sigma$ contour obtained from the six-parameter global fit, 
		restricted to the parameters $y$ and $C$ by setting the remaining 4 parameters to the values which minimise the $\chi^2$, as described in the text.}
		\label{Fig:yC}	
		\end{center}
\end{figure}

The $\chi^2$ function is relatively steep in the $y$ and $C$ directions, and depends only mildly on the other parameters.
We show in Fig.~\ref{Fig:yC} the $1\sigma$ contour plot of the six-parameter global fit as a function of $y$ and $C$. We have set the remaining four parameters to values that minimize the $\chi^2$ at each value of $y$ and $C$.
We find at the $1\sigma$ equivalent confidence level that
\begin{equation}
y = 2.06 \pm  0.05,
\label{Eq:ybound}
\end{equation}
with a corresponding value for $C$ given by $a^{3-y}C  = 6.9 \pm 1.1$.
Notice in the plot the high level of correlation between these two determinations.

\begin{figure}[h]
\begin{center}
\includegraphics[height=6.3cm]{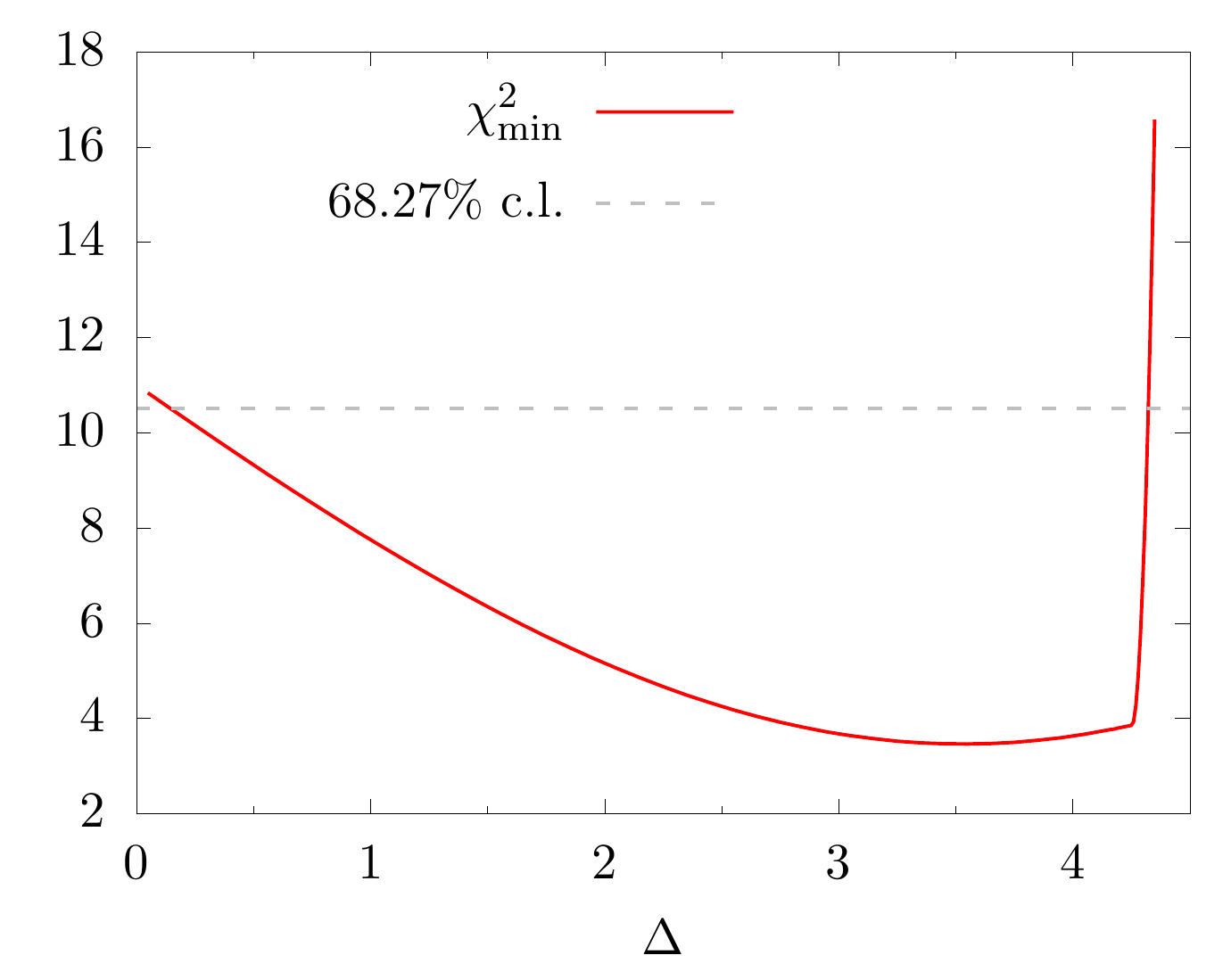}
		\caption{The chi-squared minimum as a function of $\Delta$, obtained from the six-parameter global
		 fit to the LSD data using Eqs.~(\ref{Eq:fit1})-(\ref{Eq:fit3}). The grey dashed line represents the value of the $\chi^2$
		 corresponding to the $1\sigma$ region in the full six-dimensional parameter space. In making the plot, we optimised the choice of the remaining five parameters, to minimise the $\chi^2$ for each value of $\Delta$.}
		\label{Fig:chisquarevsdelta}
		\end{center}
\end{figure}

The parameters $y$ and $C$  characterise the response of the EFT to a non-zero 
fermion mass in the underlying theory, and can be determined by fitting Eq.~(\ref{Eq:fit1}) 
alone to the (relatively accurate) lattice data for $F_{\pi}^2$ and $M_{\pi}^2$. 
This was done in Refs.~\cite{Appelquist:2017vyy,Appelquist:2017wcg}, making use of earlier LSD measurements.
Our six-parameter global fit leads to consistent results. We also repeat the exercise of performing the two-parameter fit
of $y$ and $C$ on the updated LSD measurements, and find that the central values of the fit are unaffected, but in this case $\chi^2/N_{\rm dof} = 0.26$.

The ratio $f_{\pi}^2/f_d^2$  is also relatively insensitive to the details of
the potential, in particular to $\Delta$,
but draws heavily on the measurements of $M_d^2$, 
and is hence affected by larger uncertainties. By making use of the
improved new LSD measurements we find
\begin{equation}
\frac{f^2_\pi}{f^2_d} = 0.086\pm 0.015.
\end{equation}
This determination is consistent  with the result in Ref.~\cite{Appelquist:2017vyy},
but with improved precision.

An exploration of the $\chi^2$ distribution in the full six-dimensional parameter space reveals that it is relatively flat in the $\Delta$ direction below $\Delta \sim 4.25$. In Fig.~\ref{Fig:chisquarevsdelta}, we show the result of this exploration restricted to a cut in which the other five parameters are chosen to minimise $\chi^2$ for each value of $\Delta$. The curve evolves rapidly only above $\Delta \sim 4.25$, strongly disfavoring larger values. Within the six-dimensional space, a $1\sigma$ determination imposes the restriction $\Delta \chi^2 \equiv  \chi^2 -\chi^2_\text{global min} < 7.04$ \cite{Press:2007zz}. In our case, this leads to the limit $\chi^2 \lesssim 10.4$, shown as the grey dashed line in the figure. This indicates the allowed range for $\Delta$ to be
\beqs
0.1 \lsim \Delta \lsim 4.25\,,
\label{Eq:Delta}
\eeqs
with values in the range 3 - 4 moderately preferred.  The full range also include values slightly above 4 corresponding to a ``role-reversal'' of the two terms in $V_{\Delta}$.

\begin{figure*}
\begin{center}
\begin{picture}(430,195)
\put(0,0){\includegraphics[height=6cm]{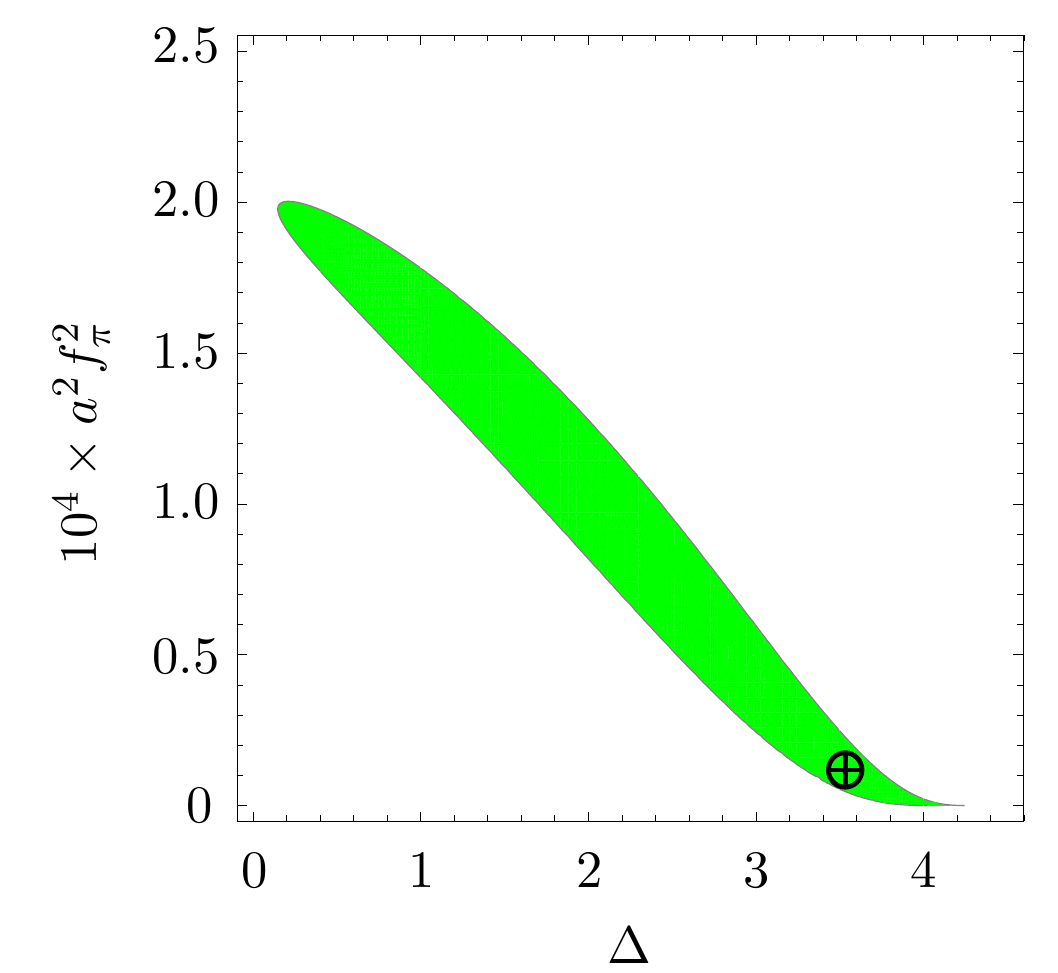}}
\put(220,0){\includegraphics[height=6cm]{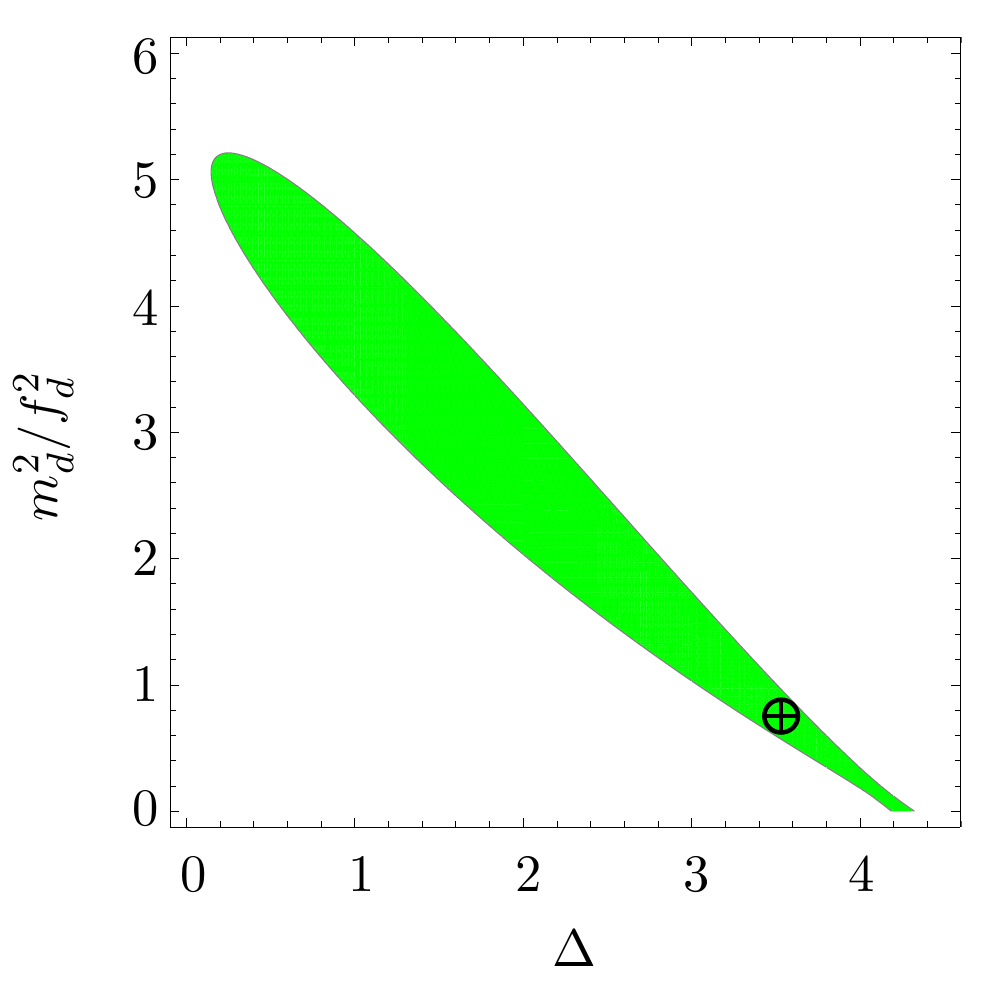}}
\end{picture}
		\caption{The $1\sigma$ region obtained using the six-parameter 
		global fit described in the main text, showing the range of $a^2f_\pi^2$ (left panel) 
		and $m_d^2/f_d^2$ (right panel) as a function of the weakly constrained parameter $\Delta$. 
		The black crosses mark the central values for the 
		fit parameters.}
		\label{Fig:fpimdvsdelta}
		\end{center}
\end{figure*}

The weakness of the constraint on $\Delta$ has to be interpreted with caution: the value of the $\chi^2$ at its global minimum is rather small, which might indicate that either the uncertainties on the input measurements are over-conservative, or that the correlations are important, or possibly both. For example, a trivial multiplicative rescaling of the global $\chi^2$ to adjust $\chi^2/N_\text{dof} =1$ at the minimum would result in restricting the allowed range to $1.6 \lsim \Delta \lsim 4.25$. Because we are taking the errors from the literature, and because no systematic study of the correlation between different measurements has been reported, we maintain our conservative result in Eq.~(\ref{Eq:Delta}) as our best estimate of $\Delta$.

The left panel in Fig.~\ref{Fig:fpimdvsdelta} is obtained by selecting values of $\Delta$ and $a^2f^2_\pi$, setting the remaining four parameters to minimise $\chi^2$ for each value of $\Delta$ and $a^2f^2_\pi$, and then shading green the corresponding points on the plot that satisfy $\Delta\chi^2<7.04$. In the right panel, we do the same, but for values of $\Delta$ and $m^2_d/f^2_d$. In this way, we indicate the extent of the six-dimensional joint $1\sigma$ confidence region in the $a^2f^2_\pi$, $m^2_d/f^2_d$ and $\Delta$ directions. The figure illustrates that the favored ranges for $a^2f^2_\pi$ and $m^2_d/f^2_d$ are correlated with $\Delta$. It also shows that if a preferred value of $\Delta$ was identified, lifting the degeneracy of the $\chi^2$ along the $\Delta$ direction, then $a^2f^2_\pi$ and $m_d^2/f_d^2$ would be determined with good precision.

The correlation between $a^2f^2_{\pi}$ and $\Delta$ is particularly informative. 
The measured values of $a^2F^2_{\pi}$ vary from up to  $3\times 10^{-3}$ for the largest fermion masses studied by the LSD collaboration, down to $4\times10^{-4}$ for the smallest ones. Its chiral extrapolation  $a^2f^2_{\pi}$ is close to this range  for the  smallest allowed values of $\Delta$,
but becomes much lower as $\Delta$ approaches $4$. For $\Delta$ near its global-minimum value $3.5$, 
$a^2f^2_{\pi}$ is one order of magnitude below currently measured values of $a^2F^2_{\pi}$. 
If future improved studies confirm that $\Delta$ lies in this range, it could pose a challenge for numerical exploration of the near-massless regime. Yet, with current precision we cannot exclude values of $\Delta$ small enough to allow improved lattice studies to reach this physically interesting regime. In Appendix \ref{Sec:appendixb}, we show a plot of $a^2F_{\pi}^2$ versus $am$ (on a logarithmic scale) for three values of $\Delta$, indicating how additional data points at smaller $am$  could begin to narrow the range of allowed $\Delta$ values.

The quantity $m_d^{2}/f_d^2$ is also strongly correlated with $\Delta$.
Its central value is ${m_d^2}/{f_d^2}\simeq 0.75$, and lies in the range $0 \lsim {m_d^2}/{f_d^2} \lsim 5.2$, throughout the entire allowed domain of $\Delta$. This shows that the self-coupling of the dilaton in the chiral limit $m_{\pi}^2 \rightarrow 0$ is relatively weak.

To summarise, the six-parameter global fit yields a low value of the $\chi^2/N_{\rm dof}$ at the minimum, validating the form of the EFT we employed.
The extraction of the parameters  $y$, $C$ and $f^2_{\pi}/f^2_d$, 
depends only weakly on the choice of potential.
Our global fit makes use of updated lattice measurements,
and is compatible with earlier determinations~\cite{Appelquist:2017vyy,Appelquist:2017wcg}.
We obtained a first measurement of $\Delta$, for which a broad range of values
$0.1\lsim\Delta\lsim 4.25$  is allowed. The determination of  the remaining two parameters is potentially more accurate, but $a^2f^2_\pi$ and $m_d^2/f_d^2$ are strongly correlated with $\Delta$. Improved measurements, especially of the mass of the scalar $M_d^2$, combined with a dedicated study of  correlations within lattice measurements, might resolve the degeneracies and identify the boundary of the chiral regime, for which $F_{\pi}^2 \simeq f_{\pi}^2$ and $M_{\pi}^2 \ll M_d^2$.

\section{Beyond leading order}
\label{Sec:Corrections}



In this section we discuss the form and magnitude of corrections to the EFT employed so far, developing the corrections as a systematic expansion in small parameters. We first discuss the EFT in the chiral limit $m=0$, starting with corrections to the dilaton potential. When only the dilaton and its self interactions are included, we find that all corrections are controlled by a single expansion parameter $m_d^2/(4\pi f_d)^2$. Then, continuing to work in the chiral limit, we include the $N_f^2 -1$ pNGBs. Counting factors depending on $N_f$ now enter the quantum loop corrections, and are accounted for.

It may be natural to postulate that the scale-breaking quantity $m_d^2/f_d^2$ depends itself on $N_f$ (relative to some critical value $N_{fc}$) in a manner emerging from the underlying gauge theory \cite{Appelquist:2010gy,Golterman:2016lsd}, but this requires moving outside the framework of the EFT. As already noted in Section \ref{Sec:Lagrangian}, we work only within the EFT. Finally in subsection \ref{Sec:beyondbeyond}, we include the explicit breaking of chiral symmetry which was employed in Sections \ref{Sec:Lagrangian} and \ref{Sec:Lattice} to fit the lattice data of the LSD group.

\subsection{The EFT in the chiral limit}
\label{Sec:Chiral}

\subsubsection*{The static potential}

In the EFT employed so far, all of the dilaton's self interactions are described by the potential $V_{\Delta}$ in Eq.~(\ref{Eq:potential}). This form includes a single scale-violating term with scaling dimension $\Delta$, and is smooth in the limit $\Delta \rightarrow 4$. Corrections to this potential $V_{\Delta}$ can be organized into a series of terms governed by the scale breaking parameter $m_d^2/f_d^2$. This can be implemented via a spurion analysis as described in Appendix \ref{Sec:appendixa}. The full potential, taken to be part of the Lagrangian density ${\cal L}$, can be expressed in the form
\begin{widetext}
\begin{equation}
V(\chi)=\label{Eq:potentialsubleading}
V_{\Delta}(\chi)+\chi^4 \sum_{n=2}^{\infty} {a}^{}_n
\left(\frac{m_d^2}{(4-\Delta) f_d^2}\right)^n\left(1-\left(\frac{f_d}{\chi}\right)^{4-\Delta}\right)^{n},
\end{equation}
\end{widetext}
where smoothness in the limit $\Delta\rightarrow 4$ is evident and where the extremum of the full potential remains at $\chi=f_d$.

This form also emerges from examining the dilaton loop corrections to $V_{\Delta}$. Since the loop-expansion parameter is naturally of order $m_d^2/(4\pi f_d)^2$, one finds that the coefficients $a_n$ are of order $1/((4\pi)^2)^{n-1}$. We develop the perturbation expansion adopting the background field method \cite{Coleman:1973jx} and employing dimensional regularization. This implements the prominent role played by scale invariance, retaining only logarithmic cutoff dependence and finite parts, disregarding (defining away) power-law cutoff dependence. The one-loop contribution to the Coleman-Weinberg effective potential takes the form

\beqs
\delta V_\text{CW}&=&
\frac{1}{64\pi^2}
\left[\left({\cal M}^2\frac{}{}\right)^2\log\left(\frac{{\cal M}^2}{\Lambda^2}\right) \,+\, c\right]\,,
\label{Eq:CW}
\eeqs
where we have replaced the pole term $2/\epsilon$ in dimensional regularization with log$(\Lambda^2)$ where $\Lambda$ is a momentum-space cutoff. Here, $c$ is a scheme-dependent constant, while ${\cal M}^2\equiv\frac{\partial^2}{\partial\chi^2}V(\chi)$. Starting at tree level we take $V=V_{\Delta}$. Extracting only the cutoff-dependent part, correcting the potential appearing in the Lagrangian density, we have 

\begin{multline}
\delta V=-\frac{1}{64\pi^2}\log\left(\frac{\Lambda^2}{\mu^2}\right)\left[\frac{m_d^2\chi^2}{(4-\Delta)f_d^2}\right]^2\\\times\left[3-(\Delta-1)\left(\frac{f_d}{\chi}\right)^{4-\Delta}\right]^2\,,
\end{multline}
where $\mu$ is a typical scale characterising the EFT, such as $f_d$. The expression $\delta V$ can be rearranged into corrections to the terms appearing in $V_{\Delta}$, supplemented by the $n=2$ term in Eq.~(\ref{Eq:potentialsubleading}). With $\Lambda$ no larger than, say, $4\pi f_d$, we can sensibly take the factor $\log\left(\Lambda^2/\mu^2\right)$ to be $O(1)$. This then leads to the estimate $a_2 \sim 1/(4\pi)^2$.

We can extend the scalar-loop expansion of the potential to higher orders, organizing the logarithmically cutoff dependent contributions at each order into the form of Eq.~(\ref{Eq:potentialsubleading}). This series contains all the zero-derivative operators required for renormalization of the potential. Each of the $a_n$ coefficients can be estimated in this way, leading, as indicated above, to $a_n\sim1/((4\pi)^2)^{n-1}$. 

Powers of $\Delta-1$ also appear in the estimates for every $a_n$. Similarly, positive powers of $\Delta-2$ appear in each $a_n$ for $n>2$ and positive powers of $\Delta-3$ appear in each $a_n$ for $n>4$. Thus, as expected, for integral values $\Delta = 1,2,3$, the scalar-loop expansion generates only a finite number of such terms. In general, the identification of $m_d^2/(4\pi f_d)^2$ as the expansion parameter for scalar-loop corrections to the potential, together with the estimate $m_d^2/f_d^2 = O(1)$ (Fig.~\ref{Fig:fpimdvsdelta}) emerging from our numerical fits, insures that Eq.~(\ref{Eq:potentialsubleading}) constitutes a systematic, controlled set of corrections to $V_{\Delta}$.

\subsubsection*{Derivative operators}

It is possible to systematically identify the tower of derivative operators via a spurion analysis, along the lines of the Appendix-\ref{Sec:appendixa} discussion of the static potential. Here we instead consider the generation of derivative operators in the loop expansion, those that appear with logarithmically cutoff-dependent coefficients.

Derivative operators, unlike the terms in $V(\chi)$, arise from loops of NGBs as well as loops of the scalar itself. The resultant operators of the effective action can involve both the scalar and NGB fields. (Loops of massless NGBs do not generate corrections to the static potential, unless equations of motion are used to recast contributions to derivative operators as contributions to the potential. We choose not to do this, and so the above results are unchanged after including NGB loops.)  

Higher derivative terms in the action correspond to corrections in momentum space with increasing powers of $p_{}^2/ (4\pi f_{\pi})^2$ and $p^2/ (4\pi f_{d})^2$, where $p$ is a characteristic momentum. With $p^2 \leq m_{d}^2$, and with the parameter values emerging from the fits of Section \ref{Sec:Lattice}, the corrections are small unless overwhelmed by pNGB counting factors.

Consider first the single-dilaton loop diagram employing $V_{\Delta}$ once and the interaction in Eq.~(\ref{Eq:Lpi}) once. Its logarithmically-cutoff-dependent contribution to the EFT Lagrangian is
\beqs
\Delta {\cal L}_{(1)}&=&  \frac{1}{64 \pi^2}\log\left(\frac{\Lambda^2}{\mu^2} \right)
\frac{f_{\pi}^2}{f_d^2} \frac{\partial^2 V_{\Delta}}{\partial \chi^2}
\Tr\left[ \partial_{\mu} \Sigma \partial^{\mu} \Sigma^{\dagger}\right] \,.
\eeqs
This two-derivative  operator describes a correction of order $m_d^2/ 2(4\pi f_d)^2$ with respect to the tree-level operator in Eq.~(\ref{Eq:Lpi}). Another logarithmically-cutoff-dependent, single-dilaton loop contribution arises from utilising the interaction in Eq.~(\ref{Eq:Lpi}) twice. It leads to the correction
\beqs
\Delta {\cal L}_{(2)}&=&  \frac{1}{64 \pi^2}\log\left(\frac{\Lambda^2}{\mu^2} \right)\frac{f_{\pi}^4}{f_d^4}
\Tr\left[\partial_{\mu} \Sigma \partial^{\mu} \Sigma^{\dagger}\right]^2  \,.
\eeqs
This four-derivative operator describes a correction to the tree-level theory of relative order $(f_{\pi}^2/f_d^2)p^2/ 2(4\pi f_{d})^2$. These examples exhibit the effective expansion parameters, which are small with the EFT-parameter values of Section \ref{Sec:Lattice}.

Corrections arising from loops of NGBs, which bring in the NGB counting factor, are typically larger. Operators involving only external scalar fields, for example, are first induced by one loop of NGBs by keeping just the quadratic-NGB-field contribution in the factor $\Tr\left[\partial_{\mu}\Sigma\partial^\mu \Sigma^{\dagger}\right]$ in Eq.~(\ref{Eq:Lpi}). Using this interaction, leads to a logarithmic-divergent, four-derivative contribution to the EFT Lagrangian given by
\beqs
\Delta {\cal L}_{(3)}&=&  \frac{1}{64 \pi^2}\log\left(\frac{\Lambda^2}{\mu^2} \right)
\frac{(N_f^2 - 1)}{\chi^2} \left(\frac{}{}\partial^{\mu}\partial_{\mu}\chi\frac{}{}\right)^2\,.\quad
\label{Eq:3}
\eeqs
The appearance of $\chi$ in the denominator of this operator, treating it as a background field in the evaluation of the NGB loop, does not indicate the presence of a dangerous singularity, as the vacuum value for $\chi$ is nonzero. 

Taking $\log(\Lambda^2/\mu^2)\sim O(1)$, this quantity makes a contribution to the dilaton two--point function of order $(N_f^2 -1)p^{2}/2(4\pi f_d)^2$ (in momentum space), relative to, say, the dilaton kinetic term.  Other four-derivative scalar-field operators generated via a one-NGB loop graph also lead to corrections of relative order $(N_f^2 -1) p^{2}/2(4 \pi f_d)^2$. Higher powers of $(N_f^2-1)$ will naturally arise at higher orders in the loop expansion, but they will be accompanied by correspondingly higher powers of $p^{2}/2(4\pi f_d)^2$.

Additional corrections from loops of NGBs arise already in familiar chiral perturbation theory. At one loop, they are of relative order $N_f p^2/2(4\pi f_{\pi})^2$, with higher powers of this quantity entering at higher orders in the loop expansion. Taken as a whole, the derivative operators describe an expansion in quantities of order $N_f^2 p^2/2(4\pi f_d)^2$ and $N_f p^2/2(4\pi f_{\pi})^2$, along with terms with fewer powers of $N_f$.

In the fits of Section \ref{Sec:Lattice} to the LSD lattice data, we have concluded that $N_{f} f_{\pi}^2/ f_d^2 \approx 0.7$. Thus the effective expansion parameters arising from the derivative operators are no larger than of order $N_f^{2}p^2/2 (4 \pi f_d)^2$. With the restriction $p^2 \leq m_d^2$, and drawing on the estimate $m_d^2/f_d^2 = O(1)$ as shown in Fig.~\ref{Fig:fpimdvsdelta}, we can conclude that the corrections to our classical EFT, as employed in the chiral limit, are likely to be no more than of order $15\%$.

\subsection{Beyond the chiral limit}
\label{Sec:beyondbeyond}

In the presence of a non-vanishing fermion mass $m$, the full tree level potential for $\chi$ is given by $W(\chi)$ in Eq.~(\ref{Eq:Wpotential}). 
The NGBs now become pNGBs, with a non-vanishing mass arising from ${\cal L}_{M}$ in Eq.~(\ref{Eq:LM}). The EFT can be recast in terms of the lattice-measured quantities 
$M_{\pi}^2$, $M_{d }^2$, $F_{\pi}^2$, and $F_{d}^2$ using Eqs.~(\ref{Eq:scaling})-(\ref{Eq:R}). We can then apply Eq.~(\ref{Eq:CW}), with $V(\chi)$ replaced by $W(\chi)$, to compute the logarithmically cut-off dependent correction to $W(\chi)$ coming from dilaton loops, beginning at the one-loop level. As in the chiral limit, these corrections are quite small even for the $N_f=8$ theory since large counting factors are not present. Dilaton-loop contributions to other quantities, described by derivative operators, are similarly small.

There are also pNGB-loop corrections to the tree-level EFT, and unlike in the chiral limit these include corrections to the static potential. They arise from the scalar-pNGB interaction present in ${\cal L}_{M}$ in Eq.~(\ref{Eq:LM}), re-expressed in terms of the capital-letter, lattice-measured quantities. The logarithmically cutoff-dependent contribution at one loop is 
\begin{multline}
\delta V(\chi)_{\rm pNGB}  =  - \frac{1}{ 64\pi^2} \log\left(\frac{\Lambda^2}{\mu^2}\right) 
\\\times(N_f^2 -1) M_{\pi}^4 \left(\frac{\chi}{F_d}\right)^{2y-4} \,, 
\label{Eq:pngb}
\end{multline}
where the exponent $2y-4$ arises from the form of Eq.~(\ref{Eq:LM}) together with the form of the pNGB kinetic 
term in Eq.~(\ref{Eq:Lpi}).
It is important to establish that such corrections, with their counting factors, 
are relatively small. 

This can be assessed by computing the effect of Eq.~(\ref{Eq:pngb})
on physical quantities such as $M_d^2$,
 determined so far by the tree-level potential $W(\chi)$. The correction is given by
\begin{multline}
\frac{\delta M_d^2}{ M_d^2}  \approx \frac{1}{ 64\pi^2} \log\left(\frac{\Lambda^2}{\mu^2}\right) (2y - 4)(2y-5)\\  
\times(N_f^2 - 1) \frac{F_{\pi}^2}{F_d^2} 
\left(\frac{M_{\pi}^2}{F_{\pi}^2}\right)  \left(\frac{M_{\pi}^2}{M_d^2}\right) \,.\label{Eq:Mdshift}
\end{multline}
Again taking $\log(\Lambda^2/\mu^2)\sim1$, using $(N_f^2-1)F_\pi^2/F_d^2\sim5.5$ as determined by the fit of Section~\ref{Sec:Lattice}, and noticing from Fig.~\ref{Fig:dataLSD} that the ratio $M^2_\pi/M_d^2$ is no larger than unity, we have $\delta M_d^2/M_d^2\lesssim5(2y-4)(2y -5)M_{\pi}^2/2(4\pi F_{\pi})^2$. Since the factor $M_\pi^2/2(4\pi F_{\pi})^2$ is no larger than 0.05, this is a small correction for $y$ of order unity. With $y$ in the range of Eq.~(\ref{Eq:ybound}) as dictated by the LSD lattice data, it is a very small correction. Similar results hold for the correction to the decay constant $F_d$.

In addition to distorting the dilaton potential in the regime away from the chiral limit, the pNGB loops also induce other operators in this regime. These include derivative operators, and additional corrections to ${\cal L}_M$ (Eq.~\ref{Eq:LM}). These terms again include pNGB counting factors and lead to various momentum-dependent effects. We have not yet compiled all these corrections to the EFT, to be included along with corrections to the static potential in the fits to the LSD lattice data. In Ref.~\cite{Appelquist:2017vyy}, we provided rough order-of magnitude estimates of these effects, concluding that some \emph{could} be relatively large for the $N_f=8$ theory. As shown more carefully above, however, contributions to these quantities arising from corrections to the static potential are relatively small.

These additional operators describe corrections that should be no larger than the distortions to the dilaton potential. Factors of $N_f$ enter together with similarly small kinematic factors. Also, it is worth noting that these factors enter in much the same way they entered the derivative operators in the chiral limit (now with measured masses and decay constants replacing the extrapolated quantities). In the static limit, as we noted in section~\ref{Sec:Chiral}, the corrections are relatively small. These observations and the fact that the tree-level EFT provides a good fit to the LSD lattice data, suggest that loop corrections are small in the regime of the data as well as in the chiral limit.

\vskip0.5 cm

While this paper was being completed the aforementioned Ref.~\cite{Cata:2019edh} became available. It discusses the same tree-level Lagrangian, and contains a study of the structure, but not the numerical size, of one-loop effects, focusing on derivative operators rather than the scalar potential.

\section{Summary and conclusions}
\label{Sec:Discussion}

We have developed and explored an EFT describing the coupling of an approximate dilaton to a set 
of pseudo-Nambu-Goldstone bosons (pNGBs) originating from the spontaneous breaking of a 
global  symmetry. We employed a tree-level scalar potential $V_{\Delta}$ in Eq.~(\ref{Eq:potential}),
 including one operator of dimension $\Delta$ responsible for the breaking of scale invariance. 
 The parameter $\Delta$ is a priori unknown. The tree-level potential is the first term in a 
  series of operators of size estimated by studying the loop expansion of the EFT. 
 The pNGBs are given a mass via a chiral-symmetry breaking operator of dimension $y$ 
 which further breaks the scale symmetry. The parameter $y$ is also unknown a priori.

We first examined the  EFT at the tree level, establishing certain scaling relations expressing lattice-measurable quantities in terms of its six free parameters. We applied these relations to the currently available lattice data for the scalar and pNGB masses $M_d^2$ and $M_{\pi}^2$ and the pNGB decay constant $F_{\pi}^2$ obtained by the LSD collaboration for the $SU(3)$ gauge theory with $N_f = 8$ fermions in the fundamental representation. We performed a maximum likelihood analysis validating the EFT interpretation of the lattice data. The parameter $y$  was confirmed to lie close to $2$, with $y = 2.06 \pm 0.05$, consistent with earlier determinations. We also obtained a first determination of the scaling dimension $\Delta$, finding that it can lie in a fairly broad range $0.1 \lesssim \Delta \lesssim 4.25$. Fig.~\ref{Fig:fpimdvsdelta} illustrates the high level of correlation between $\Delta$ and the determination of other parameters.


We then discussed corrections to the tree level analysis generated by cutoff-dependent loop diagrams within the EFT. Small ratios, breaking scale symmetry and sometimes chiral symmetry, enter all such terms. In the case of pNGB loops, there are also counting factors that can be large in the $N_f=8$ theory~\cite{Appelquist:2017vyy}. For the EFT in the chiral limit, however, we concluded, in Section~\ref{Sec:Chiral}, that with the parameter values emerging from fits to the lattice data, corrections to the tree-level potential \emph{and} corrections arising through derivative operators, are relatively small. We commented on the general form of all such corrections. 

Then, we discussed loop corrections away from the chiral limit and in the regime of the lattice data in Section \ref{Sec:beyondbeyond}. We first argued that corrections to the potential, even those arising from pNGB loops, are under control. Finally, we discussed the role of derivative operators in this regime, arguing that they should be no larger than the distortion to the potential. The full compilation of all these corrections is a future project. It would be interesting to perform this entire analysis on the data of the $SU(3)$ theory with $N_f =2$ sextets~\cite{Fodor:2012ty,Fodor:2015vwa,Fodor:2016pls,Fodor:2017nlp,Fodor:2019vmw}.

An important result of this paper is that with the present level of precision in the lattice measurements and the 
absence of a reported study of correlations, the empirical evidence allows for the breaking of scale invariance 
in the chiral limit to be clearly non-marginal. 
Within the EFT, $\Delta$ is allowed to take values well below $4$. 
As a consequence, the allowed value of $f_{\pi}^2$ ranges over more 
than an order of magnitude. If $\Delta$ is small, lattice calculations could soon reach the chiral regime, where the mass of dilaton would approach some finite value. At the extremum, this could be achievable by reducing the value of the fermion mass $m$ by only a factor of $2-3$ with respect to the smallest values already available in Ref.~\cite{Appelquist:2018yqe} (see Appendix \ref{Sec:appendixb}).

\vspace{0.0cm}
\begin{acknowledgments}

We acknowledge helpful discussions with the members of the LSD collaboration, C.-J.~D. Lin, and Y. Shamir. The work of MP has been supported in part by the STFC Consolidated Grant ST/P00055X/1.

\end{acknowledgments}

\appendix

\section{Spurion Analysis}
\label{Sec:appendixa}

We construct the EFT by starting from an exactly scale invariant theory, possessing a moduli space of degenerate vacua in which scale invariance is broken spontaneously. We then weakly deform this EFT by adding perturbations which break scale invariance explicitly. We can do this by introducing a spurion field $\lambda(x)$, which is a chiral symmetry invariant but transforms under scale transformations $x\rightarrow e^{\rho} x$ according to the rule
\begin{align}
	\lambda(x) \rightarrow e^{\rho(4-\Delta)}\lambda(e^\rho x).
\end{align}
We take the scaling dimension, $4-\Delta$, to be an unknown free parameter to be determined from lattice data.

In the chiral limit, EFT operators are built from scale and chirally invariant combinations of the fields, derivatives and the spurion $\lambda$. Scale invariance is then explicitly broken in the EFT by demoting $\lambda(x)\rightarrow\lambda$ to a constant scale. We do not expect gauge theories such as QCD, (having $N_f$ far below the conformal window boundary value) to be well described by EFTs constructed this way.

We focus here on contributions to the static potential for the dilaton. We further require it to be analytic in $\lambda$, although it is allowed to be nonanalytic in $\chi$ \cite{Rattazzi:2000hs,Goldberger:2008zz,Chacko:2012sy}\footnote{In Eq.~(23) of Ref.~\cite{Rattazzi:2000hs}, this potential was employed to provide a field-theoretical interpretation of the Goldberger-Wise stabilisation mechanism~\cite{Goldberger:1999uk} of the electroweak scale.}
\begin{align}
	V(\chi) = \chi^4 \sum_{n=0}^{\infty}b_n \left[\frac{\lambda}{\chi^{4-\Delta}}\right]^n.
	\label{eq:naivesum}
\end{align}
This potential can be re-organized into the more convenient form 
\begin{align}
V(\chi) = \chi^4\sum_{n=0}^{\infty}a_n\left[\frac{\lambda}{f_d^{4-\Delta}}\right]^n\left(1-\left(\frac{f_d}{\chi}\right)^{4-\Delta}\right)^n,
\label{eq:fullv}
\end{align}
where
\begin{align}
a_n = (-1)^n \sum_{r=n}^{\infty}b_r \left[\frac{\lambda}{f_d^{4-\Delta}}\right]^r \frac{r!}{n!(r-n)!}.
\end{align}
Note that the expression within the round brackets in Eq.~(\ref{eq:fullv}) vanishes if $\Delta=4$.
Therefore the potential (and the whole EFT) is organised as a perturbative expansion in a quantity proportional to $\lambda|4-\Delta|/f_d^{4-\Delta}$, and reliably truncated at a chosen threshold value of $n$.

In the leading order EFT, we work to linear order in $\lambda$, truncating the potential:
\begin{align}
	V_\Delta(\chi) = a_0 \chi^4 + a_1\frac{\lambda\chi^4}{f_d^{4-\Delta}}\left(1-\left(\frac{f_d}{\chi}\right)^{4-\Delta}\right)
\end{align}

We can now recast $V_\Delta$ in terms of more familiar quantities. Firstly, we can set $a_1=1/\Delta$ without loss of generality (we can always rescale $\lambda$ to compensate). We can then change to more directly physical variables
\begin{align}
	\lambda = \frac{m^2_d f^{2-\Delta}_d}{(4-\Delta)},\qquad
	a_0 = -\frac{m^2_d}{4\Delta f^2_d},\label{eq:b0def}
\end{align}
and finally recover the familiar $V_\Delta$ potential:
\begin{align}
	V_\Delta(\chi) = \frac{m^2_d}{4(4-\Delta)f^2_d}\chi^4\left[1-\frac{4}{\Delta}\left(\frac{f_d}{\chi}\right)^{4-\Delta}\right].
\end{align}

It can now be seen that the full potential is an expansion in a quantity proportional to $m_d^2/f_d^2$. As noted in Section \ref{Sec:Chiral} (Eq.~\ref{Eq:potentialsubleading}), it takes the form
\begin{multline}
	V(\chi) = V_\Delta(\chi) + \chi^4\sum_{n=2}^{\infty}a_n\left(\frac{m^2_d}{(4-\Delta)f^2_d}\right)^n\\\times\left(1-\left(\frac{f_d}{\chi}\right)^{4-\Delta}\right)^n,
\end{multline}
where we have observed in Section \ref{Sec:Chiral} that $a_n\sim 1/((4\pi)^2)^{n-1}$. This form makes it evident that the particular limit $\Delta \rightarrow 4$ is smooth for the entire potential, with $m_d^2$, $f_d^2$ and $a_n$ held fixed. In this limit, $V(\chi)$ describes corrections to the dilaton potential employed in Ref.~\cite{Golterman:2016lsd}.


\section{Approaching the Chiral Limit}
\label{Sec:appendixb}

\begin{figure}[h]
	\begin{center}
		\includegraphics[height=5.6cm]{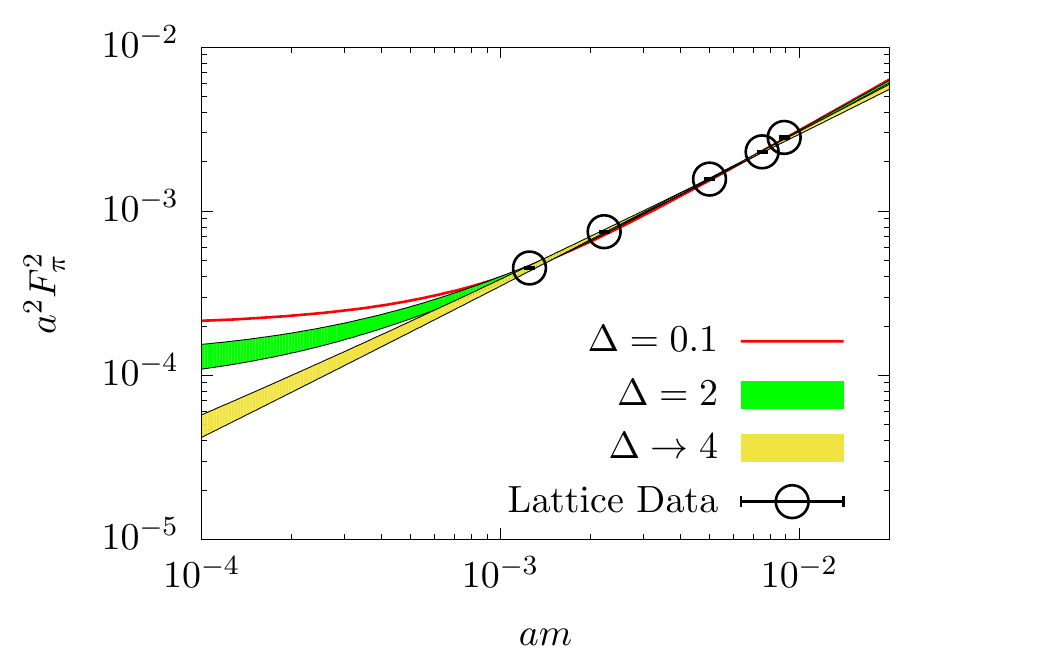}
		\caption{Figure indicating the EFT prediction for how $F^2_\pi$ varies with fermion mass as the chiral limit is approached, assuming different values for $\Delta$. More details are provided in the text.}
		\label{Fig:extrapolation}
	\end{center}
\end{figure}

To illustrate how the value for $\Delta$ affects the dependence of the pNGB decay constant $aF_\pi$ on the fermion mass $am$, we show  this dependence explicitly in Fig.~\ref{Fig:extrapolation}. Here, we compare our predictions (restricted by the parameter fits obtained in Section \ref{Sec:Lattice}) to the current LSD measurements. The black circles represent the same data points for the $N_f=8$ theory that were shown earlier in the right panel of Fig.~\ref{Fig:dataLSD}, obtained originally from Ref.~\cite{Appelquist:2018yqe}.

The widths of bands on the plot provide an indication of the uncertainties in the extrapolation of the fit results. We drew the bands and line as follows: To draw the red line, we first set $\Delta$ to 0.1, the smallest value in its allowed range. We then determined the 5 remaining model parameters by minimising the $\chi^2$. Using these values, we employed Eqs.~(\ref{Eq:Fdvariation}), (\ref{Eq:R}) and (\ref{Eq:scaling}) to draw the red line. We drew the green band by first setting $\Delta=2$ and then locating the top and bottom of the 1$\sigma$ allowed range for the parameter $a^2f^2_\pi$ (see Fig.~\ref{Fig:fpimdvsdelta}) by finding the values of $a^2f^2_\pi$ for which $\Delta\chi^2=7.04$, having minimised the $\chi^2$ with respect to the other 4 parameters. This procedure defined two sets of 6 model parameters, both of which define a line $F^2_\pi(m)$ through Eqs.~(\ref{Eq:Fdvariation}), (\ref{Eq:R}) and (\ref{Eq:scaling}). We then drew these two lines on Fig.~\ref{Fig:extrapolation} with the area between them shaded green, forming the green band. The yellow band was formed using a similar procedure, except that the limit $\Delta\rightarrow4$ was taken at the end.

The green band, for example, indicates that for $\Delta\sim2$, the extrapolated pNGB decay constant squared is only a factor of roughly $3$ below its smallest lattice-measured value.


\end{document}